\newcommand{\state} {| \phi\rangle_{\rm OUT}}   		

\newcommand{\cappa}{ {\cal K} }   		
\newcommand{\Om}{\Omega}			
\newcommand{\Oms}{\Omega_s}	
\newcommand{\Omi}{\Omega_i}	
\newcommand{\Omp}{\Omega_p}
\newcommand{\De}{ {\cal D} }
\newcommand{\OmGVM}{{\Omega_\mathrm{gvm}}}
\newcommand{\OmGVS}{{\Omega^{\prime}_\mathrm{gvs}}}
\newcommand{\OmGVSp}{{\Omega_\mathrm{gvs}}}
\newcommand{\DOmp}{{\Delta \Omega_\mathrm{p}}}
\newcommand{\tauGVM}{{\tau_\mathrm{gvm}}}
\newcommand{\tauGVS}{{\tau^{\prime}_\mathrm{gvs}}}
\newcommand{\tauGVSp}{{\tau_\mathrm{gvs}}}
\newcommand{\taup}{{\tau_\mathrm{p}}}
\newcommand{\aptilde}{\tilde{\alpha}_p}				%
\newcommand{\as}{\hat{a}_s}
\newcommand{\ai}{\hat{a}_i}
\newcommand{\aap}{\hat{a}_p}
\newcommand{\asout} {\hat{a}_s^{\mathrm{out} } }
\newcommand{\aiout} {\hat{a}_i^{\mathrm{out} } }
\newcommand{\A} {\hat{A} }
\newcommand{\Asout} {\hat{A}_s^{\mathrm{out} } }
\newcommand{\Aiout} {\hat{A}_i^{\mathrm{out} } }
\newcommand{\sinc}{{\rm sinc}}

\newcommand{\nn}{\nonumber}
\newcommand{\bsub}{\begin{subequations}}
\newcommand{\esub}{\end{subequations}}
\newcommand{\beq}{\begin{equation}}
\newcommand{\eeq}{\end{equation}}
\newcommand{\beqa}{\begin{eqnarray}}
\newcommand{\eeqa}{\end{eqnarray}}
\newcommand{\beql}{\begin{subequations}\begin{eqnarray}}
\newcommand{\eeql}{\end{eqnarray}\end{subequations}}
\documentclass[pra,aps,amsmath, nofootinbib]{revtex4}
\usepackage{amsmath} 
\usepackage{graphicx}
\usepackage{float}   
\usepackage{verbatim}  
\usepackage{braket}
\usepackage{color}
\begin{document}
\title{ Temporal coherence and correlation of counterpropagating twin photons}
\author{Alessandra Gatti$^{1,2}$,Tommaso  Corti$^2$, Enrico Brambilla$^2$, }
\affiliation{$^1$ Istituto di Fotonica e Nanotecnologie del CNR, Piazza Leonardo  Da Vinci 32, Milano, Italy;  $^2$  
Dipartimento di Scienza e Alta Tecnologia dell' Universit\`a dell'Insubria, Via Valleggio 11,  Como, Italy}
\begin{abstract}
This work  analyses the temporal coherence and correlation of counterpropagating twin photons  generated in a quasi-phase matched nonlinear cristal by spontaneous parametric-down conversion.  We find out  different pictures depending on the pump pulse duration relative to  two characteristic temporal scales, determined  respectively by  the temporal separation between the counter-propagating and the co-propagating wavepackets.  When the pump duration is intermediate between the two scales, we show  a transition from a highly entangled state to an almost separable state, with strongly asymmetric spectral properties of the  photons.  
\end{abstract}
\maketitle
\section*{Introduction}
\label{sec:intro}
Spontaneous parametric down-conversion  (SPDC) 
occurring in $\chi^{(2)}$  media 
 is one of the most accessible  sources  both of entangled photon pairs and of single photons, heralded by detection of  the partner. The microscopic process, where a high energy photon of the pump laser splits into two lower energy photons,  is ruled by  conservation laws (energy,  momentum,  angular momentum, polarization), which are at the origin  of a wide range of  quantum correlations between the members of the pair. 
\par
In standard co-propagating configurations,  the two-photon state  is characterized by a {\em high dimensional entanglement}, 
because a quantum correlation is present  over   huge  temporal and angular bandwidths. The temporal correlation  was historically the first one to be studied \cite{Hong1987}:  a down-conversion event can take place anywhere along the crystal, so that the arrival time of the twins is not known. However, 
the members of a pair,   generated at the same point,   propagate nearly in the same direction, and exit the crystal almost simoultaneously
 A small uncertainty in their temporal  separation is present because of their different group velocities (type II) or because of the  group velocity dipersion (typeI), and can  be reduced to its smallest limit  (the optical cycle) when the spatial degrees of freedom are properly controlled \cite{Gatti2009},\cite{Ottavia2012}. Such a short correlation time results in a high-dimensional temporal entanglement \cite{Mikhailova2008}. Its spectral counterpart is  the huge spectral bandwidth of SPCD emission, and the high dimensional spectral entanglement of SPDC  photons\cite{Avenhaus2009}.  High-dimensional entanglement  offers  relevant opportunities in view of broadband quantum communication schemes,  but can also be regarded as a negative feature, because it affects the purity of heralded single photons. 
\par
\begin{figure}[h,b,t]
\begin{center}
\includegraphics[scale=0.55]{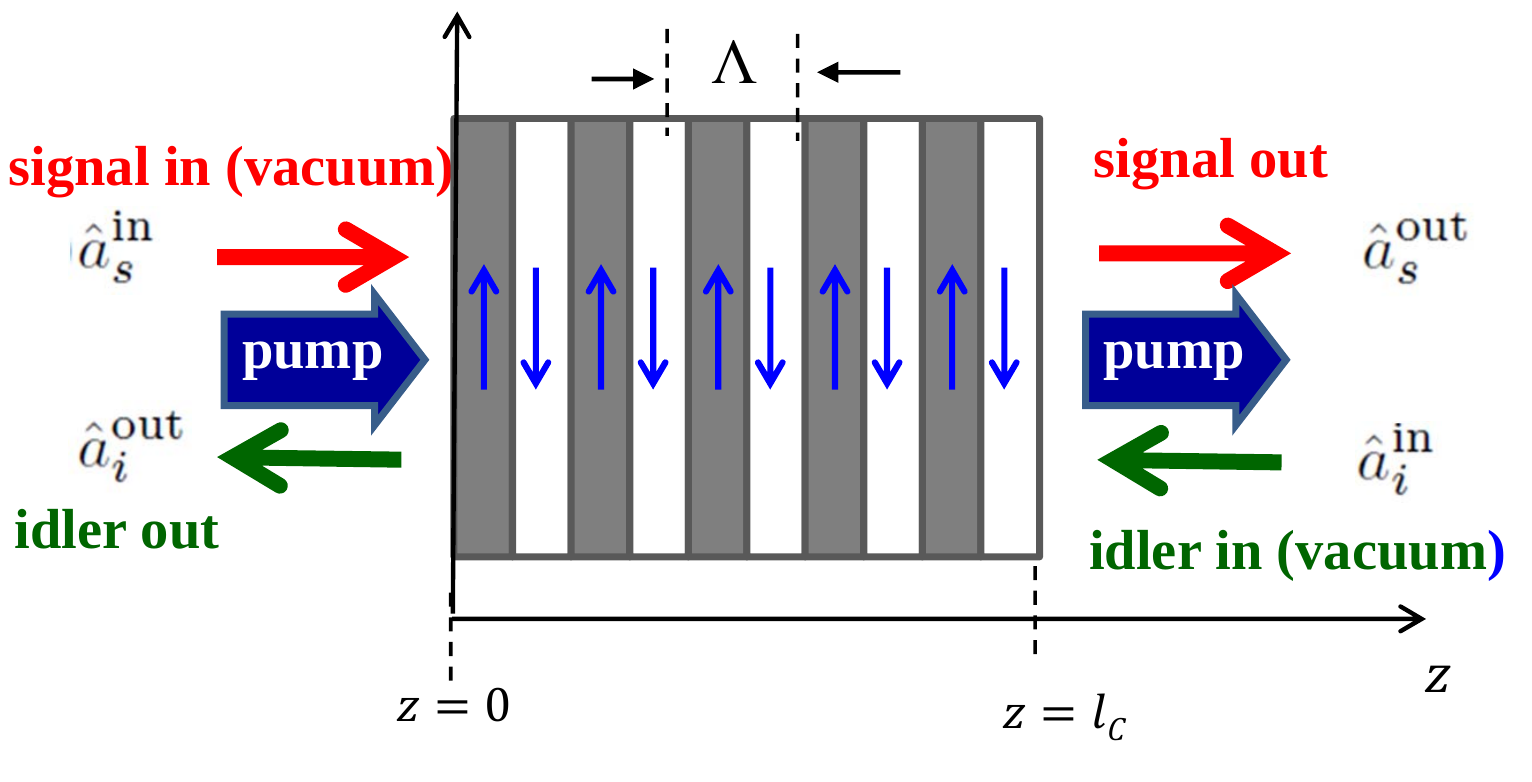}
\caption{(Color online) Geometry of the counterpropagating down-conversion (see text).   }
\label{scheme}
\end{center}
\end{figure}

This work considers a  non-conventional configuration, where one of the  down-converted  photons is  generated in the backward direction with respect to the pump laser, in a  periodically poled crystal  (Fig.\ref{scheme}).
Although predicted almost fifty years ago \cite{Harris1966}, counter-propagating down-conversion has been only recently realized \cite{Canalias2007,Stromvqist2012}, thanks to technical advancements in achieving the  sub-micrometer poling periods necessary to phase-match the interaction \cite{Pasiskevicius2015} . 
\par
Counterpropagating PDC presents unique features, as the presence of a threshold for the pump intensity beyond which coherent parametric oscillations take place \cite{Ding1996}, thereby the name {\em Mirrorless Optical parametric Oscillator}  (MOPO)\cite{Canalias2007}. 
 In a related work \cite{Corti2015},  we study 
the quantum correlation of counter-propagating twin  beams close to the  threshold; here,  instead,  we focus on the regime of spontaneous photon pairs production, well below threshold, and  analyse the  temporal quantum properties  of counterpropagating  twin photons generated in a purely collinear configuration.  
\par
A second  peculiar feature of the MOPO
 is the narrow spectral  bandwidth of  emission (the backward propagating wave can be even more monochromatic than the pump laser \cite{Canalias2007}).   In the quantum domain, 
as pointed out in \cite{Christ2009},   counterpropagating SPDC can   generate highly monochromatic photon pairs in an 
almost separable state,  which makes it a promising source of  high-purity heralded single photons.  
\par
In this work we  provide 
 a detailed theoretical analysis  of the  effects of the spectral properties  of the pump laser on the degree of entanglement of the state, identify the physical conditions under which the state may become separable, and provide a consistent interpretation of the transition from an entangled to a separable state. 

 In particular we will show that the system dynamics is governed by two well separated time scales: a {\em long one},  related to   the temporal separation of counterpropagating waves, which is on the order of the  transit time of light along the crystal (tens of picoseconds), and a 
{\em short} one related to the temporal separation of  co-propagating waves due to their different group velocities (order 1ps or smaller).  When the pump pulse duration is  intermediate between the two scales,  we will show that the two-photon state  becomes separable, and remains separable for a wide range of pump durations, 
whilst it has a high degree of entanglement in the two opposite limits.   Notice that  such a difference of  time scales occurs naturally in  the counterpropagating configuration,  for basically any kind of material and tuning condition.  This is quite different from 
the co-propagating case where separability of the state requires special operational points\cite{Mosley2008}.    
\par
In addition, we shall investigate the coherence properties of the SPDC photons taken individually,   showing a transition from a symmetric state, for a long pump pulse, to a highly asymmetric state for a short pump pulse. In particular, in the regime where  the state  is separable, the spectrum of the signal turns out to reproduce the spectrum  of the co-propagating pump laser, while that of the backward propagating idler is entirely determined by the crystal properties. 

The paper is organized as follows: Section \ref{sec:model} introduces a  quantum model for counterpropagating PDC; Sec. \ref{sec:psi} characterizes the quantum correlation  of twin photons in  the spectral domain, while Sec.\ref{sec:coherence} is devoted to their coherence properties. Sec.\ref{sec:Schmidt} quantifies the degree of entanglement of the state via the Schmidt number. Finally Sec.\ref{sec:temporal} gives  an interpretation of the transition from entanglement to separability by analysing the correlation in the temporal domain. 
\section{The model } 
\label{sec:model}
The starting point of our analysis are the equations describing  the propagation of three 
 interacting pump,  signal and idler waves along  a slab of periodically poled second order nonlinear crystal (Fig.1).   We shall consider  here only collinear propagation,    either assuming that light is collected only at  small propagation  angles with respect to the pump,  or because of a waveguiding configuration.
 In a crystal with periodic inversion of the nonlinear susceptibility, the momentum conservation in the three-wave interaction is  replaced by a less restrictive  conservation law (quasi-phase matching)\cite{Boyd} , which includes also  the momenta
$
 \frac{2 \pi}  {\Lambda} m
$
of the nonlinear lattice, where $\Lambda$ is the poling period (for example $m= \pm 1, \pm 3...$ for a simple poling).  Since the effective nonlinearity is higher for lower orders, one usually tries to phase-match the first order $m=\pm 1$ interaction. 
The counter-propagating configuration, in which  one wave  (say the idler)  is generated in the backward  direction with respect to the pump laser (Fig.1),  thus requires a  poling period on the  order of the pump wavelength, because the pump momentum needs to be almost entirely compensated by  the grating momentum $ k_{G}= \frac{2 \pi}  {\Lambda} $.   In these conditions, 
 the central frequencies   $\omega_s$,  $\omega_i=\omega_p -\omega_s$  of the down-converted  wavepackets are determined by  quasi-phase matching  at the central pump frequency $\omega_p$ 
\beq
k_{0s} -k_{0i}= k_{0p} -  \frac{2 \pi}  {\Lambda} 
\label{Delta0}
\eeq
where 
 $k_{0j}  =n_j (\omega_j) \omega_j /c$, 
$j=s,i,p$ are the wave numbers at the three central frequencies. We shall mostly focus on the commonly realized  type O interaction \cite{Canalias2007}, where the three waves have the same polarization, but  we leave the formalism quite general. Hence  the subscript j in the wave number may  refer to dispersion relations for either the ordinary or extraordinary wave, including thus  type II or I PDC.  
\par
We than introduce the positive frequency part of  field operators (with dimension of photon destruction operators) for the three wavepackets as 
\bsub
\label{adef}
\beqa
\hat{A}_s (\Om, z) &=& e^{ +i k_s (\Om) z }    \as(\Om, z) \, , \\
\hat{A}_i (\Om, z) &=& e^{ -i k_i (\Om) z }   \ai(\Om, z) \, , \\
\hat{A}_p (\Om, z) &=& e^{ +i k_p (\Om) z }   \aap(\Om, z) \, , 
\eeqa
\esub
where capital $\Omega$ is a frequency offset  from the carrier frequencies,  and $k_j (\Om)$  are the wave numbers at frequency $\omega_j + \Om$.  In this definition, the factors  
$ e^{ \pm i k_j (\Om) z } $
account  for  all the  effects of the linear propagation along the medium.  Hence the operators  $\hat{a}_j$  have a slow variation along the crystal  because  they evolve  only under  the effects of the nonlinear interaction. Their coupled equations of propagation can thus be written as ( see \cite{Corti2015} for a more detailed analysis)
\bsub
\label{prop}
\begin{align}
\frac{\partial }{\partial z} \as  (\Om, z)  &= \chi \int d \Om ' \aap (\Om + \Om ',z ) \ai^\dagger (\Om ', z) 
e^{- i \De (\Om, \Om ') z}  \, , \\
\frac{\partial }{\partial z} \ai  (\Om, z)  &=  -\chi \int d \Om ' \aap (\Om + \Om ',z ) \as^\dagger (\Om ', z) 
e^{- i \De (\Om', \Om ) z }\, , \\
\frac{\partial }{\partial z} \aap  (\Om, z)  &= - \chi \int  d \Om ' \as (\Om ', z) \ai (\Om - \Om ',z ) 
e^{ i \De (\Om, \Om-\Om ') z } \, , 
\end{align}
\esub 
where $\chi$ is proportional to the effective second order susceptibility of the  crystal,  and  only the first order terms  $\pm 1$  in the Fourier espansion of the periodic nonlinear suceptibility  have been retained   (namely order -1 for signal and idler, order +1 for the pump) . In these equations 
\beq
\label{Delta}
\De (\Om, \Om') = k_s (\Om) -k_i (\Om') -k_p(\Om + \Om ') '+  \frac{2 \pi}  {\Lambda}
\eeq
is the effective phase mismatch that rules the efficiency of each elementary down-conversion process, where   a signal and an idler photon of   frequencies $\omega_s +\Om$, $\omega_i + \Om'$ are generated out of a pump photon of frequency 
$\omega_p +\Om+ \Om'$.  Notice that, a part from the different form of the phase matching \eqref{Delta} , the only formal difference  with  the usual co-propagating case (see e.g. \cite{Brambilla2013})  is the minus sign appearing at r.h.s of \eqref{prop}b  for the counterpropagating idler. As we shall see in the following, however, this minus sign leads to  very relevant physical differences. 
\subsection{Low-gain limit}
In a parent work \cite{Corti2015} we analyse these equations in a generic gain regime, including also the region close to the MOPO threshold. In this work we instead focus on the ultra-low gain regime, much below the MOPO threshold, where photons pairs are  generated by purely spontaneous down-conversion. In this regime, the depletion of the pump beam can be neglected and the pump approximated by a constant  c-number field, corresponding to the pump pulse at the crystal input face
\beq
\aap (\Om, z) \to \alpha_p(\Om, z) \approx  \alpha_p (\Om, z=0) 
\eeq
The strength of the parametric coupling is then described by 
 the dimensionless gain parameter 
\beq
g= \sqrt{2\pi} \chi \alpha_p (t=0)  l_c \, ,
\eeq
where $\alpha_p (t=0) $ is the peak value of the pump temporal profile. 
Notice that in the limit of a monochromatic pump \cite{Suhara2010,Corti2015}  $g=\pi/2$ represents the threshold for the MOPO.  Conversely, in the limit $g \ll 1$  Eqs.\eqref{prop} can be solved perturbatively.  Namely, we write the formal solution of \eqref{prop}, starting from  the  boundary conditions: 
\beqa
\as (\Om, z=0 ) &=&  \as^{\mathrm{in} } (\Om  ) \\
\ai (\Om, z=l_c ) &=&  \ai^{\mathrm{in} } (\Om  )
\eeqa 
determined by the input signal and idler fields,  entering the crystal  from the left face of at $z=0$ 
and from the right face at $z=l_c$, respectively  (Fig.1) .  By solving iteratively,   a perturbative  seriers of powers of $g $  is obtained. By keeping only the first order terms in $g \ll 1$ ,  one obtains a Boguliobov  linear trasformation that links the output to the input operators: 
\bsub 
\label{inout}
\beqa
\asout (\Om_s)  &=&  \as (\Om_s, z=l_c )  \\
                               &=&  \as^{\mathrm{in} } (\Om_s  ) + \int d \Om_i  \psi (\Om_s, \Om_i)  
                                                                                                                             \ai^{\mathrm{in} \dagger } (\Om_i  )\, , 
\label{inout1} \\
\aiout (\Om_i)  &=&  \ai (\Om_s, z=0 )  \\
                               &=&  \ai^{\mathrm{in} } (\Om_i  ) + \int d \Om_s  \psi (\Om_s, \Om_i)  
                                                                                                                             \as^{\mathrm{in} \dagger } (\Om_s  )
\label{inout2} \, .
\eeqa
\esub
with the {\em  biphoton amplitude} given by 
\beqa
\label{psi}
\psi (\Om_s, \Om_i) &=&  \frac{g }{ \sqrt{2\pi} } \aptilde(\Om_s + \Om_i)  \nn \\
					& \times&  \sinc \left[      \frac{ \De (\Om_s, \Om_i)l_c}{2}  \right] 
					 e^{ -i  \frac{\De (\Om_s, \Om_i) l_c}{2} }
\eeqa
where 
\beq
\aptilde (\Om)  = \int \frac{d t}{ \sqrt{2\pi} } e^{i \Om t} \frac{ \alpha (t)  }{\alpha_p(t=0)   }
\eeq
is the  Fourier profile of the pump pulse at the crystal input face, normalized to its temporal  peak value. 
Notice that Eqs.\eqref{inout} define a unitary tranformation only up to first order in $g$. 
In the following,  the input signal and idler field at the left and right end faces of the crystal will be taken in the vacuum state.

It is worth remarking that  the quantum field formalism here employed can be  replaced by an equivalent state formalism (see e.g \cite{Gatti2012} for more details) where the state evolves instead of the field operators. By applying the trasformation \eqref{inout} to the input vacuum state,  one obtains at the output the usual state
\beq 
\state = \left |0 \right \rangle + \frac{1}{2} \int d\Om_s d\Om_i \psi (\Om_s, \Om_i) \as^\dagger (\Om_s) \ai ^\dagger (\Om_i) \left |0\right \rangle \, , 
\label{state}
\eeq
describing  the superposition of the  vacuum state $\left |0 \right \rangle$  and of a two-photon state,  
where  the  photon pair can   be  generated in any of the  Fourier modes  $\Oms, \Omi$ with probability amplitude $\psi(\Oms, \Omi)$. 
  In this respect,  the formalism used here is equivalent to the one employed in \cite{Christ2009,Booth2002}.
\section{Spectral biphoton correlation}
\label{sec:psi}
This section is devoted to the analysis of the biphotonic   correlation  in the spectral domain. Precisely,  we focus on  the probability amplitude  $\langle\Asout (\Om_s) \Aiout(\Om_i)  \rangle$  of finding  a pair of photons at frequencies $\Om_s, \Om_i $ at the crystal output faces. Using   the input-output relations \eqref{inout} and the definitions \eqref{adef}: 
\beq
\langle\Asout (\Om_s) \Aiout(\Om_i)  \rangle = e^{i k_s (\Om_s) l_c }\psi (\Om_s, \Om_i) \, . 
\eeq
with $\psi$  given by Eq.\eqref{psi}. 
As usual, the biphoton correlation is   the product of two terms:  i) the pump spectral amplitude $\aptilde(\Om_s + \Om_i)$,  reflecting  the energy conservation in the microscopic process,  and ii) the phase matching function $\sinc ( \De l_c/2)  
e^{-i \De l_c/2} $,   reflecting  the generalized momentum conservation. Concerning the latter, we can expand  $\De(\Om_s, \Om_i)$  in Eq.\eqref{Delta}  in power series 
of the frequency shifts from the carriers.  Down-conversion spectra are typically  narrow    \cite{Canalias2007,Stromvqist2012}, as will become also clear in the following, so that one is allowed to retain only  terms  up to first order 
\beqa
\De (\Om_s, \Om_i)\frac{l_c}{2}     &\approx& \frac{l_c}{2} \left[ ( k'_s -k'_p)  \Om_s - ( k'_i + k'_p)  \Om_i  \right]  \\
						&=&  -\left (\frac{ \Om_s}{\OmGVM}+  \frac { \Om_i } {\OmGVS} \right) \, ,
\label{Delta1}
\eeqa
where the zero order term vanishes because of Eq.\eqref{Delta0}, and 
$k'_i = \left. dk_j/d\omega  \right|_ {\omega=\omega_j} $ , $j=s,i,p$.
 We thus see the appearence of the two characteristic temporal scales:  
\beqa
 \tauGVM&:=\OmGVM^{-1} = \frac{1}{2} \left[ \frac{l_c}{v_{gp}} -\frac{l_c}{v_{gs}}  \right] \label{GVM}   \\
\tauGVS& :=\OmGVS^{-1}  = \frac{1}{2} \left[ \frac{l_c}{v_{gp}} +\frac{l_c}{v_{gi}}      \right]  \label{GVS}  \, , 
\eeqa
where $v_{gi} = 1/k'_i$  are the group velocities of the three wavepackets at the central frequencies.  The first  scale  [Eq.\eqref {GVM}]describes the " small" temporal  separation  between the co-propagating waves due to  their  group velocity mismatch (GVM) . The second one [Eq.\eqref {GVS}]  accounts for the "large"temporal separation of the counter-propagating pump and idler waves, which  is ruled by   the time needed by the pulse centers  to cross  the crystal.  Closely related, 
\beq
\tauGVSp  = \OmGVSp ^{-1} =  \frac{1}{2} \left[\frac{l_c}{v_{gs}} +\frac{l_c}{v_{gi}}   \right]    \label{GVSp}  \, ,    
\eeq
 describes the characteristic temporal separation  between the arrival times of an idler and a signal photon at their exit faces.  
 Clearly, since group velocities are close,  $\tauGVSp \approx \tauGVS$,  while  $\tauGVM \ll \tauGVS, \tauGVSp$, and 
\beq
\eta= \frac{\tauGVM} {\tauGVS}=\frac{\OmGVS} {\OmGVM} \ll 1  \, . 
\eeq
Therefore, the phase matching has  two well separated scales of variation: as a function of the signal frequency it decays on the broad bandwidth  $\OmGVM$, while as a function of the idler frequency  it decays on the narrow bandwidth  $\OmGVS$. 
\begin{figure}[h,b,t]
\begin{center}
\includegraphics[scale=0.65]{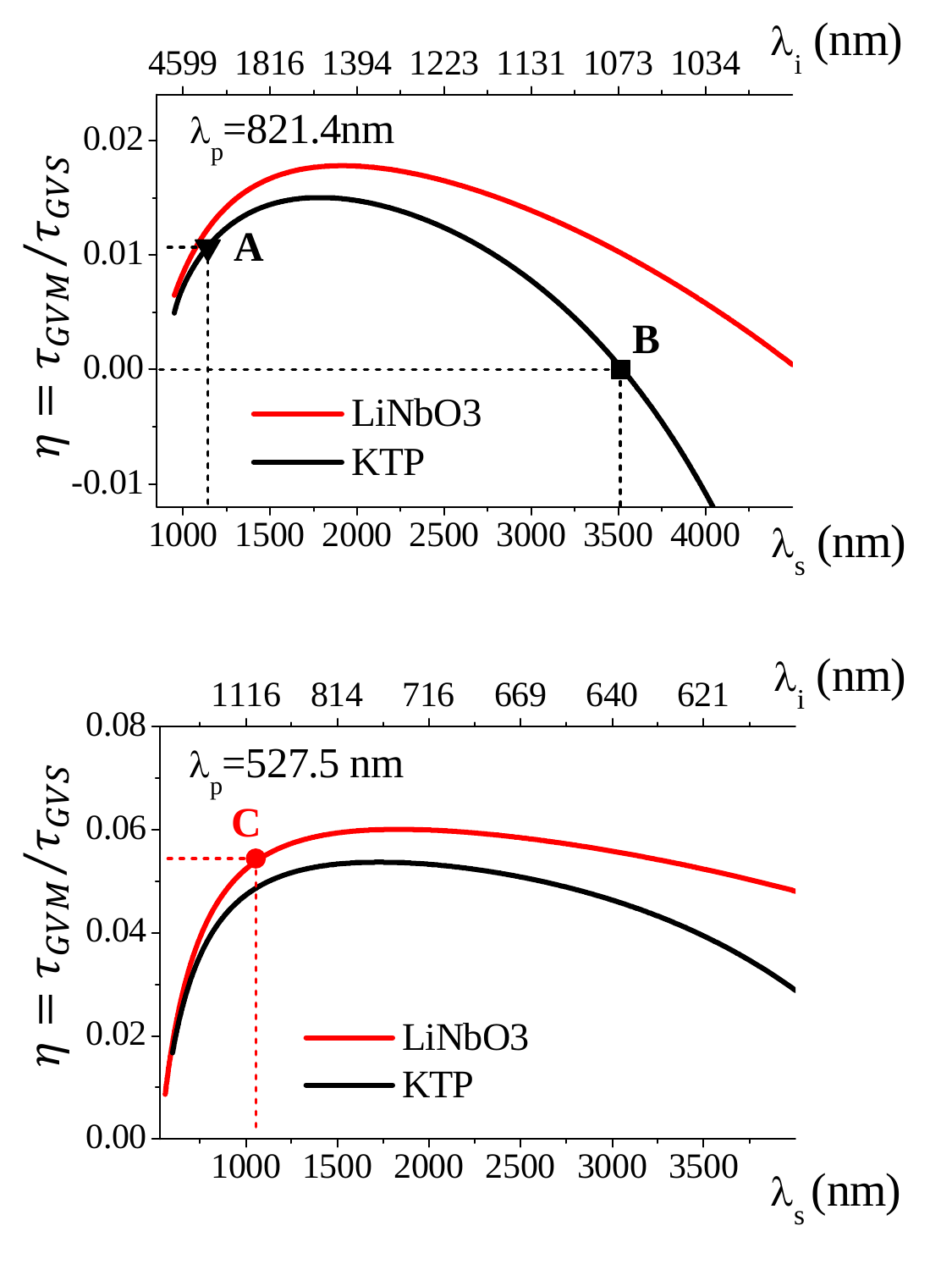}
\caption{(Color online) Ratio $\eta=\tauGVM / \tauGVS$  for periodically poled  KTP 
and   LiNbO$_3$, pumped in the infrared or visible, for type 0 $e \to ee$  down-conversion. Point {\bf A} is  KTP pumped at $\lambda_p=821$nm, with $\Lambda_{pol}=800$nm,   $\lambda_s=1141$nm, 
$\lambda_i=2932$nm, corresponding  to the experiment in \cite{Canalias2007};  
{\bf B}  is the zero GVM point for the KTP at $\lambda_p=821$nm, corresponding to  $\Lambda_{pol}=290$nm,   $\lambda_s=3523$nm, 
$\lambda_i=1071$nm. 
 {\bf C} is a LiNbO$_{3} $ slab pumped at $\lambda_p=527.5$nm,  for degenerate PDC at $\lambda_s= \lambda_i= 1055$nm, 
with $\Lambda_{pol}=236$nm.  }
\label{eta}
\end{center}
\end{figure}
Plots of the parameter $\eta$, for  periodically poled KTP (potassium titanyl phosphate)
and  LiNbO$_3$  
(lithium niobate), are shown in Fig.\ref{eta},  where  {\bf A,B,C}  are the points that will be used as examples in the following.
\par
Finally,  a third relevant scale is  the pump  spectral bandwidth. For a coherent Gaussian pump 
$\alpha_p (t) = \alpha_p(0) \exp{-\frac{t^2}{2 \tau_p^2} }$, 
the pulse duration $\tau_p$ is the inverse of the bandwidth 
\beq
\taup= \frac{1}{\DOmp }
\eeq
\begin{figure*}[h,b,t]
\begin{center}
\includegraphics[scale=0.7]{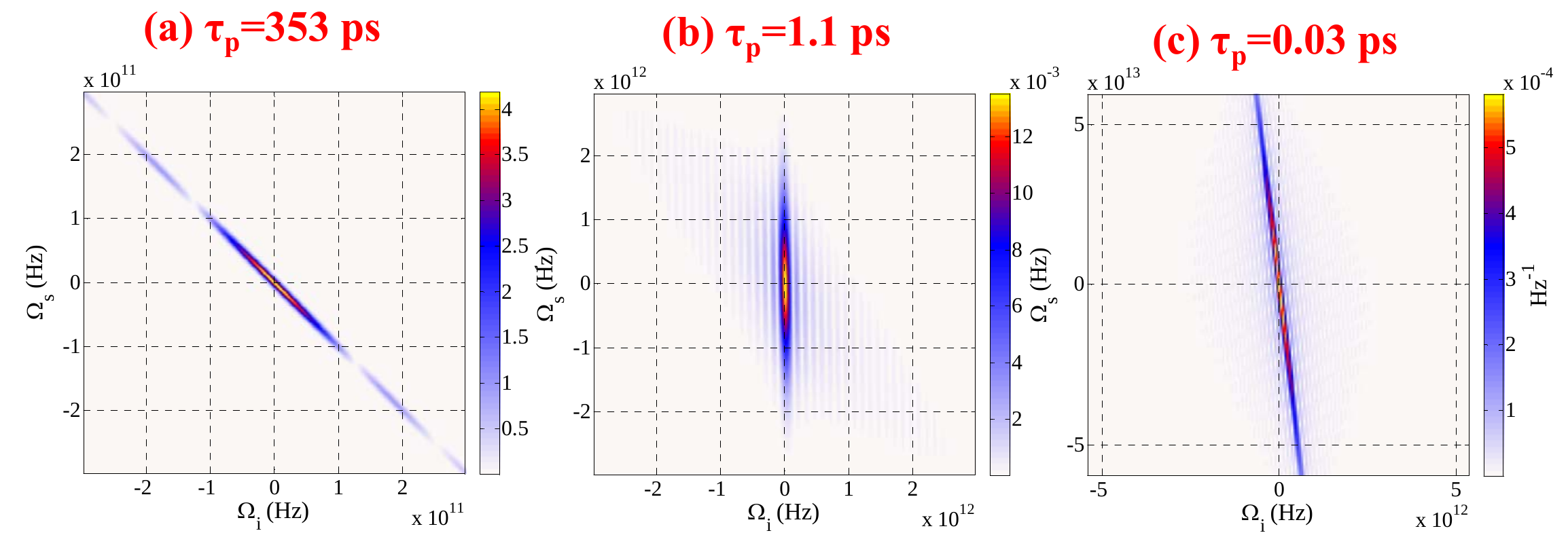}
\caption{(Color online) Biphoton correlation $\left|\psi \right|$ [Eq.\eqref{psi}]  in the plane $(\Om_i, \Om_s)$ , in various pumping regimes. Example  of a 4mm PPKTP, pumped at 821.4nm, corresponding to the point {\bf A} in Fig.\ref{eta}, with 
$\tauGVS = 25.2$ ps, $\tauGVM= 0.27 $ ps. 
 (a) Quasi CW  pump pulse  $\taup = 253 $ps. (b) Intermediate pump pulse $\tau_p= 1.1 $ ps.  (c) Ultrashort pump  $\taup= 0.03$ ps .  Note the different scales of the plots $10^{11} \to 10^{13}$ Hz. } 
\label{fig2}
\end{center}
\end{figure*}
Depending on the pump bandwidth relative to the  spectral scales of phase matching,   different physical situations arise. The three relevant possibilities, depicted in Fig.\ref{fig2},  will be  studied separately in the following. \\
%

 \noindent {\bf i) Limit of a CW pump: }   \par
We assume a narrowband pump pulse, such that
\beq  
\tau_p \gg \tauGVS \gg \tauGVM ,  \;  {\rm or} \;   \DOmp \ll \OmGVS \ll \OmGVM \, .
\label{L1}
\eeq
This limit  corresponds to a  pump pulse that in the $z$ direction  is much longer than the crystal slab, and for 
 a crystal of some mm length requires  a  pulse duration of  hundreds of picoseconds or longer. 
In this limit
the  pump spectral profile $\aptilde (\Om_s +  \Om_i) $  is much narrower than the phase matching bandwidths,   and
 the geometry   of the correlation is   dominated  by energy conservation, which requires that the twins are generated  at  symmetric frequencies 
$\Oms +\Omi =\Omp \approx 0$.  As a consequence,  the biphoton correlation \eqref{psi}  has a sharp maximum along the diagonal $\Om_s = -\Om_i$,  as shown by Fig.\ref{fig2}a. 
Indeed,  as derived  in the Appendix \ref{A1}, in this limit   the  correlation is well  approximated by : 
\begin{align}
\psi(\Oms, \Omi) &\simeq \frac{g }{ \sqrt{2\pi} } \aptilde(\Om_s + \Om_i) 
\sinc \left(     \frac{\Oms}{\OmGVSp}    \right) e^{-i \frac{\Oms}{\OmGVSp} } 
 \label{CWpsi1} \\
 &\simeq\frac{g}{ \sqrt{2\pi} } \aptilde(\Om_s + \Om_i) 
\sinc  \left( \frac{\Omi}{\OmGVSp} \right)     e^{i \frac{\Omi}{\OmGVSp} } 
 \label{CWpsi2} 
\end{align}

\par \noindent 
 {\bf ii) Limit of an ultrashort pump pulse:} 
\par
We consider here the limit: 
\beq
 \tau_p \ll \tauGVM, \tauGVS   ,  \;  {\rm or} \;      \DOmp \gg \OmGVM, \OmGVS    \,  ,
\label{L2}
\eeq 
where the  pump pulse is not only shorter than the crystal length, but also shorter than  the average separation between the pump and signal wavepackets due their GVM.  In our examples this corresponds to duration  shorter than  100 fs.  In these conditions,   the pump spectral profile $\aptilde (\Om_s+  \Om_i) $ decays slowly with respect to   $ \sinc{\De (\Om_s, \Om_i)l_c/2}$, so that   the geometry of the biphoton correlation is dominated  by  the phase matching, i.e. by the momentum conservation. As a result 
(Fig.\ref{fig2}c)  the biphoton correlation takes the  approximated form
\begin{align}
\psi(\Oms, \Omi)  &\simeq  \frac{g }{ \sqrt{2\pi} } \aptilde\left[\Om_s (1-\eta) \right]  
 \sinc   \left(     \frac{\Oms}{\OmGVM}   +   \frac{\Omi}{\OmGVS} \right)    
    e^{i     \left(     \frac{\Oms}{\OmGVM}   +   \frac{\Omi}{\OmGVS} \right) }     
 \label{ultrashortpsi1} \\
 & \simeq \frac{g }{ \sqrt{2\pi} } \aptilde\left[-\Om_i \frac{1-\eta}{\eta}\right]  
\sinc   \left(     \frac{\Oms}{\OmGVM}   +   \frac{\Omi}{\OmGVS} \right)
 e^{i     \left(     \frac{\Oms}{\OmGVM}   +   \frac{\Omi}{\OmGVS} \right)      } 
 \label{ultrashortpsi2} 
\end{align}
When plotted in  the plane $(\Om_i, \Om_s)$, the function  shows  a sharp maximum along the line  \beq
\Om_s = -\Om_i \frac{\OmGVM}{\OmGVS}
\label{line}
\eeq
where phase matching occurs (see  Eq.\eqref{Delta1}), and very asymmetric spectral properties of the signal -idler photons.\\

\par \noindent
{\bf iii) Intermediate pump pulse:} 
\par
The intermediate case, where
\beq 
  \tauGVS \gg \tau_p \gg \tauGVM  ,  \;  {\rm or} \;   \OmGVS \ll  \DOmp  \ll  \OmGVM \,  ,
\label{L3}
\eeq
 is the most peculiar one, because the biphoton correlation may approach a separable function of $\Omega_s, \Om_i$ (Fig.\ref{fig2}b).
 First of all, we remark  that  the limit \eqref{L3}   is strictly realized only for   $\eta=\tauGVM / \tauGVS \to 0 $, i.e for a vanishing  group velocity mismatch between the pump and the signal. This condition is favourable to separabilty, because  as  $ \eta \to 0$ the phase matching function tends to become a stripe  parallel  to  the  $\Oms$ axis (see Eq. \eqref{line} ), 
but it is not a sufficient one,  because of the role of  the pump profile in Eq.\eqref{psi}.  
 However, provided that  the  the pump spectrum satisfies 
 the intermediate  limit \eqref{L3},  it can be   demonstrated (Appendix \ref{A1}) 
that the  the biphoton amplitude  \eqref{psi} approaches the factorized form: 
\beq
 \psi (\Oms, \Omi) \to  \frac{g }{ \sqrt{2\pi} }     
\aptilde (\Oms)  e^{i \frac{\Oms}{\OmGVM}  }   
\times 
\sinc \left(  \frac{\Omi}{\OmGVSp}  \right)   e^{i  \frac{\Omi}{\OmGVS}  }    \,  ,  \label{intermediatepsi} 
\eeq
 i.e. it  becomes the  product of a function of $\Oms$, reproducing the pump profile, and a function of $\Omi$,  
corresponding to the phase matching profile. This describes  a nonentangled biphoton state, with the signal photon generated  in the same spectro- temporal mode as the pump, while the spectral mode of the idler is  dictated by the phase matching "$\sinc$" function of width  $\OmGVSp$.   
\\
This qualitative picture will be confirmed by the evaluation of the Schmidt number in Sec. \ref{sec:Schmidt}, and  will be further interpreted and discussed in the light of the temporal correlation of biphotons described in 
Sec.\ref{sec:temporal}
\section{Spectral coherence of counterpropagating photons}
\label{sec:coherence}
This section is devoted to the marginal statistics   of individual  signal and idler photons. The focus is  on their  spectral coherence properties, studied by means of the first order coherence functions
\begin{align}
  G_s^{(1)} (\Om, \Om')       &=     e^{-i [k_s (\Om')-k_s(\Om) ] l_c } 
\langle \A^{\dagger \, {\mathrm out} }_s  (\Om) 
\A^{{\mathrm out} }_s (\Om') \rangle
 \nn \\
 G_i^{(1)} (\Om, \Om') 
&=   \langle \A^{\dagger \, {\mathrm out} }_i  (\Om) 
\A^{{\mathrm out} }_i  (\Om') \rangle    
\label{G1idef}
\end{align}
(where a propagation  phase  factor is present in the  first definition  just for convenience of notation).
From the input-output relations \eqref{inout} one has: 
\begin{align}
 G_s^{(1)} (\Oms, \Oms') &= \int d\Omi  \psi^* (\Om_s, \Om_i) \psi (\Om_s', \Om_i) 
\, , \label{G1s} \\
G_i^{(1)} (\Omi, \Omi') &=  
 \int d\Oms  \psi^* (\Om_s, \Om_i) \psi (\Om_s, \Om_i') 
\, .  \label{G1i} 
\end{align}
i.e. the coherence functions are convolution integrals over the biphoton amplitude  $\psi$,  given by  Eq\eqref{psi} .
The knowledge of the $G_j^{(1)}$ is sufficient to determine all the statistical properties of the marginal distributions.   
For example,  the autocorrelation  of the light intensities $\hat{I}_j =\hat{A}_j^\dagger \hat{A}_j$ is given by 
\beq
\langle \hat{I}_j (\Om) \hat{I}_j (\Om') \rangle = \delta (\Om-\Om') \langle \hat{I}_j (\Om) \rangle + 
\langle \hat{I}_j (\Om) \rangle \langle \hat{I}_j (\Om') \rangle  
 +\left| G_j^{(1)} (\Om, \Om') \right|^2\,  , 
\label{Siegert}
\eeq
where $\langle \hat{I}_j (\Om) \rangle = G_j^{(1)} (\Om, \Om)$. 
This relation, which is  a consequence of the factorization theorem of Gaussian moments,  is typical of thermal-like statistics. As a matter of fact, the marginal distributions of the output signal-idler light are thermal-like Gaussian, when there is vacuum at the input. In the  low-gain regime considered here, the dominant term is the first one, i.e. the "shot-noise" term $\delta$-correlated in frequencies.  Therefore, as well known  in this regime the statistics of photon counts in each arm is Poissonian. \\
 On the other side,  the convolution integrals in Eqs.  \eqref{G1s}, \eqref{G1i} indicate  that an autocorrelation of spectral fluctuations inside each individual signal or idler wave exists because of  second order processes, that involve the probability amplitudes of generating at two pairs of  photons. 

\begin{figure*}[h,b,t]
\begin{center}
\includegraphics[scale=0.6]{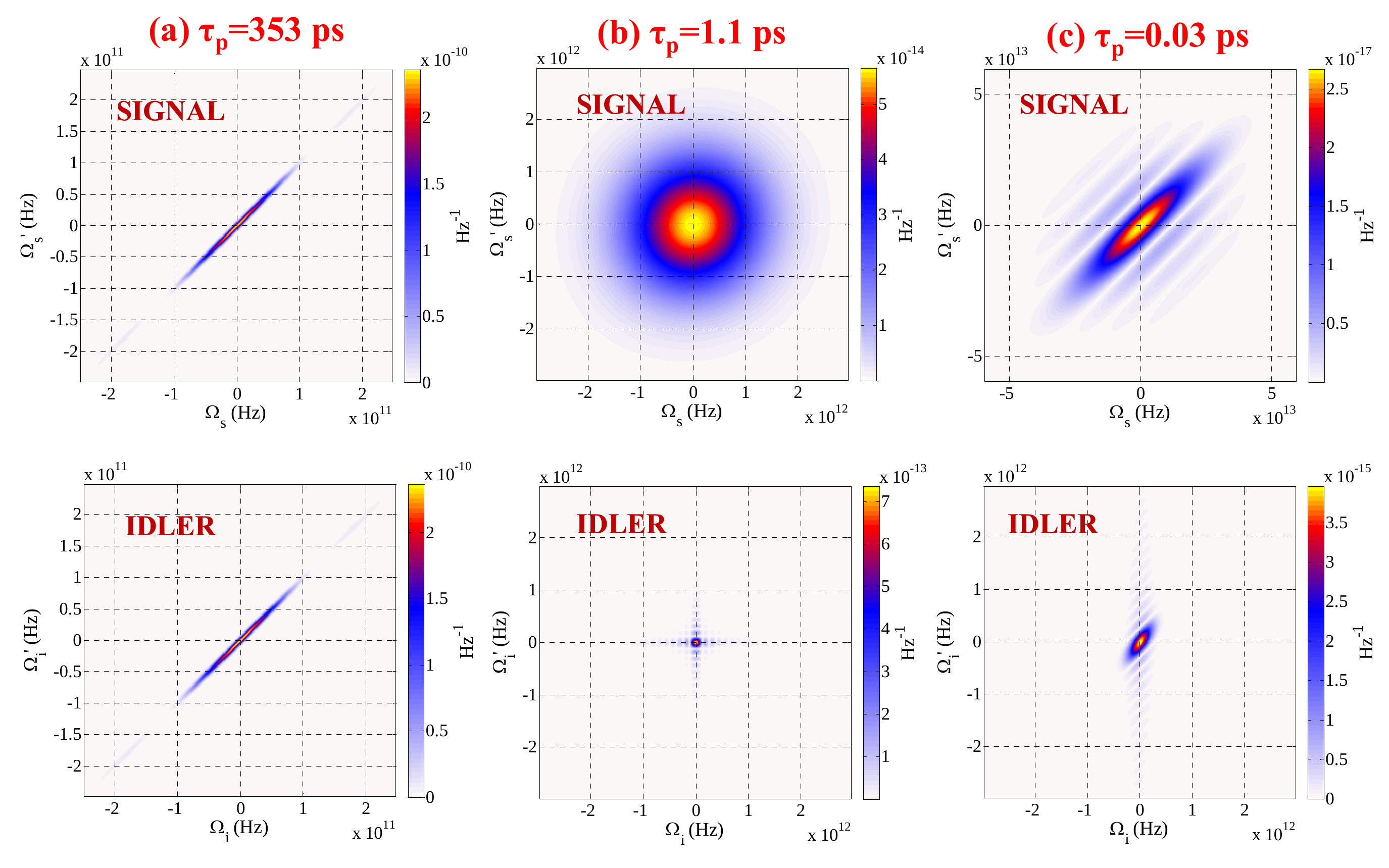}
\caption{(Color online) The coherence functions  $\left|G (\Om, \Om')\right|$   of the forward signal and backward propagating idler are plotted in the  upper and lower  row, respectively,   for different pumping regime . Column (a) Quasi CW  pump pulse  $\taup = 353 $ps. (b) Intermediate pump pulse $\tau_p= 1.1 $ ps.  (c) Ultrashort pump  $\taup= 0.03$ ps.  Same KTP crystal  slab as in Fig.\ref{fig2} (point {\bf A}  in Fig. \ref{eta}), with  $\tauGVS = 25.2$ ps, $\tauGVM= 0.27 $ ps.  Note the different scales in the panels
 }
\label{G1_all}
\end{center}
\end{figure*}

In the following we shall illustrate the three relevant cases.  The coherence functions will be evaluated  both  numerically (Fig.\ref{G1_all})  and analytically.  In the first case,  the complete Sellmeier relations \cite{boeuf2000} will be used to compute the integrals  in \eqref{G1s}, \eqref{G1i}, while the linear approximation for phase matching will be exploited to derive  approximated analytical formulas. \\
%

 \noindent {\bf i) Limit of a CW pump: }   \par
Column (a) of  Fig.\ref{G1_all} shows an example of  the signal and idler  coherence functions in the plane  $(\Om, \Om') $, numerically computed in the case of 
 a long pump pulse $\tau_p \simeq 14 \tauGVS$.  \\
In the  limit $\tau_p \gg \tauGVS$,  approximated expressions for the coherence functions can be calculated  by inserting the formulas for the biphoton correlation \eqref{CWpsi1} and  \eqref{CWpsi2}, valid in this limit, into  Eqs. \eqref{G1s} and \eqref{G1i}, respectively, and performing the simple integrals. After some passages:
\begin{align}
 G_i^{(1)} (\Om, \Om') & \simeq    G_s^{(1)} (\Om, \Om')    \\
&   \stackrel{  {\taup} \gg {\tauGVS} }   { \longrightarrow} 
\tilde{\cal I}_p(\Om' - \Om ) \,  g^2 \sinc^2 \left( \frac{\Om}{\OmGVSp} \right)  
\label{G1CW}
\end{align}
where 
$
\tilde{\cal I}_p( \Om ) = \int \frac{d t}{ {2\pi} } e^{i \Om t}\left| { \alpha_p (t)  }/{\alpha_p(0)   }\right|^2 
$
is the Fourier transform of the pump intensity profile.  These  approximated formulas have been checked  with the numerical  results and show an excellent match. These results may be considered the more refined  version of the much simpler CW model analysed in \cite{Suhara2010, Corti2015},  with  the narrow peak 
$\tilde{\cal I}_p( \Om'-\Om )$ being  the finite counterpart of the singular Dirac $\delta $ appearing in the strictly CW pump model \cite{Corti2015}.  
\\
For a quasi-CW pump the  counter-propagating signal and idler photons are predicted to have  identical spectral coherence properties. In particular,   by looking at  the $G^{(1)}$ functions along the diagonal $\Om'=\Om$  we see that their spectra $\langle \hat{I}_j (\Om) \rangle = \langle \hat{A}_j^\dagger  (\Om) \hat{A}_j (\Om)\rangle $
\begin{align}
\langle \hat{I}_s (\Om) \rangle = \langle \hat{I}_i (\Om) \rangle  
\simeq
 \frac{g^2 \tau_p }{\sqrt{2\pi}}
\sinc^2 \left( \frac{\Om}{\OmGVSp} \right) \, ,
\label{spectra_CW}
\end{align}
are identical and entirely determined by the narrow bandwidth of phase matching $\OmGVSp$ . This bandwidth is in turn the inverse of the characteristic separation $\tauGVSp$ 
between the arrival time  of an idler and a signal photon at their crystal end faces, which  rouhgly corresponds to the {\em long}  transit time of light along the crystal slab, because they propagate in opposite directions. 
 As already noticed in \cite{Suhara2010}, and as will be further discussed in Sec\ref{sec:temporal} this is clearly a big difference with the copropagating case. There,  the temporal uncertainty between the arrival times  of  the idler and signal photon is short, because determined at most by the group velocity dispersion or mismatch, which results in the huge down-conversion  bandwidths that characterize the standard co-propagating configuration. \\
On the other side,  when studied as a function of $ \Om'-\Om$ the   $G^{(1)}$  gives the characteristic size of spectral fluctuations, i.e. the spectral {\em coherence length}. This  is 
 determined by the pump bandwidth, more precisely by the width $  \sqrt{2} \Delta \Om_p$ of $\tilde{\cal I}_p (\Om'-\Om)$,  which is  much narrower than the spectral bandwidths $\OmGVSp$.  We can heuristically estimate the number of modes by counting the number of coherence length contained in the spectrum: therefore, for such a long pulse   we expect each signal and idler photon to be generated in a highly incoherent and multimode state, with the number of modes 
$ \propto \frac {\OmGVSp}{\Delta \Omp} = \frac {\taup}{\tauGVSp}$. \\

 \noindent {\bf ii) Ultrashort pump pulse: }   \par
When the pump pulse shorten below the transit time $\tauGVS$ along the crystal slab, the spectral properties of the counterpropagating idler and signal change drastically,  becoming strongly asymmetric.  First we consider the case of an ultrashort pulse,   $\taup \ll \tauGVM$  (i.e. such that pump and the signal tend to split apart during propagation). 
 The asymmetry between the forward and backward propagating photons   can be clearly appreciated  in  the third column of Fig. \ref{G1_all}, 
which plots their coherence functions
for    $\tau_p \approx 0.1 \tauGVM$. \\
Approximated expressions for the coherence functions are derived also in this case,  by using the limit   behaviour of  the biphoton correlation described by  Eqs. \eqref{ultrashortpsi1} and \eqref{ultrashortpsi2}. With some calculations: 
\beq
  G_s^{(1)} (\Om, \Om')   \stackrel{  {\taup} \ll {\tauGVM} }   { \longrightarrow}  \frac{g^2 \OmGVS}{2}  
  \left|\aptilde [\Om (1-\eta) ] \right|^2   
\sinc \left( \frac{\Om' -\Om} {\OmGVM} \right)  
 e^{-i \left( \frac{\Om' -\Om} {\OmGVM} \right) } 
\label{G1sultrashort}
\eeq
This formula predicts that the spectrum of the forward propagating signal
\beq
\langle \hat{I}_s (\Om) \rangle = \frac{g^2 \OmGVS}{2}  
  \left|\aptilde [\Om (1-\eta) ] \right|^2 
\label{signal_spectrum_ultra}
\eeq
 is a replica of   the pump spectrum 
with a scale factor $\frac{1}{1-\eta} =\frac{k'_p + k'_i }{k'_i + k'_s}    $  on the order unity.  The coherence length of the signal (the characteristic size of spectral fluctuations) is instead determined by the width of the narrower sinc function, ${l}_{coh,s} \approx \OmGVM$.   From this picture we thus expect that the signal photon, when detected independently from its twin,  is in a incoherent multimode state, with the number of modes $\propto \frac{\Delta \Omp}{ (1-\eta) \OmGVM} $ . \\
In a similar way, for the idler photon we  get: 
\beq
\ G_i^{(1)} (\Om, \Om')     \stackrel{  {\taup} \ll {\tauGVM} }   { \longrightarrow}  \frac{g^2 \OmGVM}{2}  
  \left|\aptilde [- \Om \frac{1-\eta}{\eta}  ] \right|^2  
\times \sinc \left( \frac{\Om' -\Om} {\OmGVS} \right)
 e^{-i \left( \frac{\Om' -\Om} {\OmGVS} \right) } 
\label{G1iultrashort}
\eeq
This formula predicts an idler bandwidth much narrower than the pump, precisely it predicts that  the idler spectrum follows the pump spectrum with a scale factor $\frac{\eta}{1-\eta} =\frac{k'_p - k'_s}{k'_i + k'_s}   \ll 1  $ .  The coherence length of the idler is  ${l}_{coh,i} \approx \OmGVS$,  so that  the number of temporal modes  is predicted to scale as  
$\frac{\eta \Delta \Omp}{ (1-\eta) \OmGVS} = \frac{\Delta \Omp}{ (1-\eta) \OmGVM} $ , which  is the same number as for the signal (as it must be because the signal and idler are the two members of the same entangled state, and their reduced states must exhibit the same Schmidt dimensionality, see next section) .\\
Notice that this  particular scaling of the bandwidths  of  the forward and backward propagating waves  with the pump  bandwidth 
 is well known in the literature concerning the  MOPO. There,  the same scaling factors,  $ \frac{k'_p + k'_i }{k'_i + k'_s} $ for the forward-propagating signal and $\frac{k'_p - k'_s}{k'_i + k'_s} $  for the backward propagating idler,  
 are predicted  to occurr \cite{Canalias2007, Stromvqist2012}, by using  arguments based on the phase-matching characteristic of the process.  Here, however, the analysis concerns the quantum properties of the single photons generated  well below the MOPO threshold. Moreover, at difference with the classical analysis in \cite{Canalias2007}, such  a scaling with the pump spectrum is predicted  only in rather extreme conditions, corresponding 
to an ultrashort pump pulse  $\tau_p \ll \tauGVM$.
Notice that this limit imposes a precise and not trivial constraint  on the minimum observable bandwidth of the idler photon:  
 the behaviour described by  Eq.\eqref{G1iultrashort} is indeed realized only for $\tau_p \ll \tauGVM$ , or for $\DOmp \gg \OmGVM$, 
so that it requires that 
the idler bandwidth 
\beq
\delta \Omega_i \simeq    \frac{\eta}{1-\eta}  \DOmp \gg   \frac{\eta}{1-\eta} \OmGVM = \OmGVSp
\label{constraint}
\eeq

 \noindent {\bf iii) Intermediate pump pulse: }   \par
When $ \tauGVM \ll \tau_p \ll \tauGVS$,  the properties of the twin photons are actually intermediate between the two former cases, with the forward propagating signal photon replicating  the  pump spectrum, while the coherence properties of the backward propagating idler are determined by phase matching.  
 These features are clearly exhibited by  the central column (b) of Fig.\ref{G1_all},  which  plots  a numerically computed  example of  the  coherence functions for   $\tau_p =  0.04 \tauGVS  \approx 4 \tauGVM$,  short with respect to the transit time along the slab, but long enough that GVM does not play a relevant role. \\
The observed features are a straightforward  consequence of  the separable form  \eqref{intermediatepsi}  of the biphoton amplitude 
which holds in this limit. Indeed, by using  Eq.\eqref{intermediatepsi},  in the limit $ \tau_p / \tauGVS \to 0 \; , {\tauGVM/\taup \to 0 }   $  we obtain: 
 \begin{flalign}
G_s ^{(1) } (\Om, \Om')  \to &  \frac{g^2 \OmGVSp }{ 2 }     \aptilde^* \left[ \Om \left( 1  -\eta \right) \right]   
 \aptilde \left[ \Om' \left( 1  -\eta \right) \right]     \label{G1sintermediate}  \\
G_i ^{(1) } (\Om, \Om')  \to  &  \frac{g^2  \taup}{ \sqrt{2\pi} }   \sinc \left( \frac{\Om}{\OmGVSp}	\right) 
\sinc \left( \frac{\Om'}{\OmGVSp}	\right)  e^{i \frac{\Om' -\Om}{\OmGVS}}   
 \label{G1iintermediate} 
\end{flalign}
Thus in this case the signal spectrum is a replica of the broad   pump spectrum 
$I_s(\Om) \propto  \left| \aptilde \left[ \Om \left( 1  -\eta \right) \right]   \right|^2$ , while the idler spectrum is determined by the much narrower phase-matching function 
$ I_i (\Om) \propto \sinc^2 \left( \frac{\Om}{\OmGVSp}\right) $. 
Precisely, the signal spectrum is described by the same formula \eqref{signal_spectrum_ultra} as in  the ultrashort pump case, while the idler spectral properties are described by the same formula \eqref{spectra_CW} that hoòlds in the CW pump limit. 
However, notice that in the present case the coherence properties are remarkably different, as  the two coherence functions are perfectly symmetrical along the two diagonals $\Om \pm \Om'$: as can be easily inferred from Eqs. \eqref{G1sintermediate} and \eqref{G1iintermediate} the two  coherence lengths  are   $l_{coh,s} \approx \DOmp$  and   $l_{coh,i} \approx \OmGVSp$, i.e. they are equal to the respective spectral widths. This is in accordance with the separability of the biphoton state, which corresponds to single-mode, almost coherent  reduced states for each of the two twin photon taken separately.

We conclude this section observing that the results \eqref{spectra_CW}, \eqref{constraint} and \eqref{G1iintermediate} implies  that in any pumping regime  the idler bandwidth cannot be narrower than the phase matching bandwidth $\OmGVSp$,  a limitation that arises from the imperfect momentum conservation due to the finite length of the crystal slab. 
\section{Schmidt number of entanglement}
\label{sec:Schmidt}
So far our considerations about the number of modes and the degree of entanglement of the system 
 have been qualitative.  A quantitative measure of the entanglement is offered by  the so-called
Schmidt number \cite{ekert1994,parker2000}, which  is recognized to give an estimate of the number of Schmidt modes partecipating in the entangled state, i.e. of the effective dimensionality of the entanglement \cite{exter2006}. 
First of all, as usual, we consider  the state conditioned to a photon count 
\beq 
|\phi _{\mathrm C}\rangle =  \int d\Om_s d\Om_i \psi (\Om_s, \Om_i) \as^\dagger (\Om_s) \ai ^\dagger (\Om_i) \left |0\right \rangle \, , 
\label{statec}
\eeq
where with respect to the true output state \eqref{state}, the vacuum term has been dropped. Then, we introduce  the Schmidt number, 
as the inverse of the purity of the state of each separate subsystem
\beq
\cappa= \frac{1}{ {\mathrm Tr} \{\rho_s^2\} }= \frac{1}{ {\mathrm Tr} \{\rho_i^2\} }
\eeq
where $\rho_s$,  $\rho_i$  are the reduced density matrix of the signal  and  idler , e.g. $\rho_s= {\mathrm Tr}_i \{  |\phi _{\mathrm C}\rangle \, \langle 
\phi _{\mathrm C}|  \} $. 
For a two-particle state of the form \eqref{statec}, 
 the Schmidt number can be calculated via an integral formula, as e.g derived  in \cite{Gatti2012}   
  (see also \cite{Mikhailova2008}) ,
\beq
\cappa = \frac{ {\cal N}^2} {B}   
\label{kintegral}
\eeq
where
\begin{align}
 {\cal N}  = & 
\int d\Om   \, G_s^{(1) }(\Om, \Om)     = \int d\Om   \, G_i^{(1) }(\Om, \Om) 
\label{enne} \\
B= & \int d\Om \int  d\Om' \left|G_s^{(1) }(\Om, \Om')\right|^2 
=  \int d\Om \int  d\Om' \left|G_i^{(1) }(\Om, \Om')\right|^2     
\label{B} 
\end{align}
As can be easily checked, ${\cal N}$ is 
 the espectation value (first order moment)  of the photon number  operator
$
\hat{N}_j= \int d\Om  \hat{I}_j (\Om)  
$
 in either  the  signal or idler  arm
\beq
 {\cal N}=\langle \hat{N_s} \rangle= \langle \hat{N_i} \rangle
\eeq
The quantity at denominator
is instead linked  to the  second order moment of the photon number.  By performing the integral of Eq.\eqref{Siegert} over the two spectral arguments, one gets:
\beq
B=  \langle: \hat{N}_j^2: \rangle - \langle \hat{N}_j\rangle^2  \qquad (j=s,i)
\label{B3}
\eeq
where the symbol 
 $ : \: :  $ indicates normal ordering.  
In terms of the  normalized $g^{(2)}$  coefficient: 
\beq
g^{(2)} = \frac { \langle: \hat{N}_j^2: \rangle }{\langle \hat{N}_j\rangle^2} = 1 + \frac{1}{\cappa}
\label{g2}
\eeq
In this way, as recognized in \cite{Laiho2011, Christ2011}, the Schmidt number can  be related to measurable   statistical  properties of light. In particular,  formula \eqref{g2} is well know to describe the statistics of multi-mode thermal light, with $\cappa$ playing the  role of  the ''degeneracy factor"  characterizing  the effective number of independent modes in a thermal beam.
\par
\begin{figure}[h,t]
\begin{center}
\includegraphics[scale=0.7]{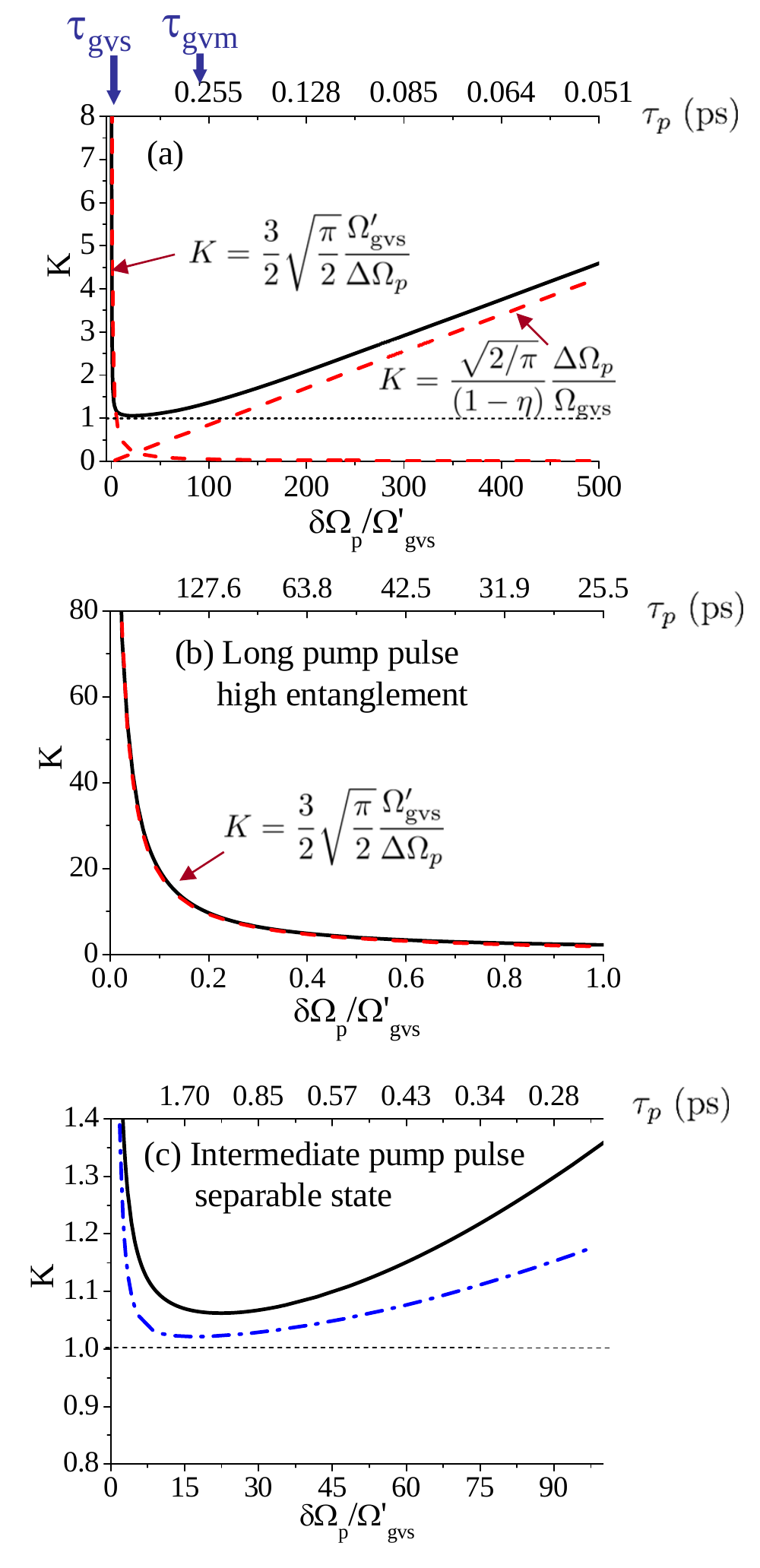}
\caption{ Schmidt number, as a function of the pump spectral  bandwidth (lower axis) or  duration (upper axis). (b) and (c) are insets of  (a), showing the transition from  high  entanglement for a long pump  $\tau_p \gg \tauGVS$ 
 to an almost separable state for $\tauGVS \gg \tau_p \gg\tauGVM$. The red dashed lines in (a) and (b) are the calculated  asymptotic  behaviours, the blu dash-dot line in (c) is the result of a Gaussian approximation. 
 4 mm PPKTP {\bf A} in Fig.\ref{eta},  with $\tauGVS=25.5 $ ps $\tauGVM=0.27$ps, $\eta=0.01$,  other parameters as in Fig.\ref{fig2}}
\label{kappa_001}
\end{center}
\end{figure}

 
Figure \ref{kappa_001} shows our results for the Schmidt number. The solid lines plot  the "exact" results, where $\cappa$ has been calculated by numerically performing the integrals involved in  \eqref{enne}, \eqref{B}, with the phase matching calculated via the complete Sellmeier relations. 
The red dashed lines in plot (a) and (b) are asymptotic behaviours, analytically derived 
by exploiting the  linear approximation for phase matching.   In particular, by using the approximated formula \eqref{G1CW} for the coherence function, and performing the integrals involved in \eqref{enne} and \eqref{B}, one
 obtains the limit of the Schmidt number for a long pump pulse 
\beq
  \cappa  \stackrel { \tau_p \gg \tauGVS} {\longrightarrow}
\frac{3}{2}\sqrt{\frac{\pi}{2}}
\frac{\OmGVS}{\Delta\Omega_p} 
\label{kappaCW}
\eeq
For an ultrashort pump pulse, the asymptotic  behaviour of $\cappa$  is calculated by using  formula \eqref{G1iultrashort} or \eqref{G1sultrashort},  for either the  signal 
or the idler  coherence function (identical results are indeed obtained). In this case
\beq
 \cappa \stackrel { \tau_p \ll \tauGVM} {\longrightarrow}  \frac{1}{1-\eta}\sqrt{\frac{2}{\pi}}
\frac{\Delta \Omp}{\OmGVM}
\label{kappaultrashort}
\eeq
The calculated asymptotes are well in accordance with our qualitative estimates of the number of modes in Sec.\ref{sec:coherence},  based on the ratio between the spectral bandwidth and the coherence length. 
\par
This shape of the curve,  showing a minimum of $\cappa$ for a given  value of the  pump bandwidth and linear asymptotes at small and large values of the bandwidth,  is commonplace,  with a qualitatively similar curve characterizing  also the co-propagating case in either  temporal \cite{Mikhailova2008} or spatial \cite{Law2004} or even spatio-temporal \cite{Gatti2012} domains. 
 The novelty here is that the minimum value of $\cappa$ is very close to unity, and remains close to unity for a rather large range of $\Delta \Omp$ (see panel (c) in Fig.\ref{kappa_001}). 
This represents indeed a big difference compared to the  copropagating case, where in order to generate separable biphotons  very special matching conditions have to be chosen, corresponding to a zero group velocity mismatch between the pump and one of the twin photons, which can be realized only in type II interactions \cite{Grice2001,Mosley2008}.

In the backward propagating case the conditions for separability are very easily approached, and rely entirely on the fact that $\eta=\tauGVM/\tauGVS$ is naturally a very small quantity,  because  the temporal separations$\tauGVM, \tauGVS$  between the co-propagating and the counterpropagating waves are on well separated time scales. 

\begin{figure}[h,t]
\begin{center}
\includegraphics[scale=0.7]{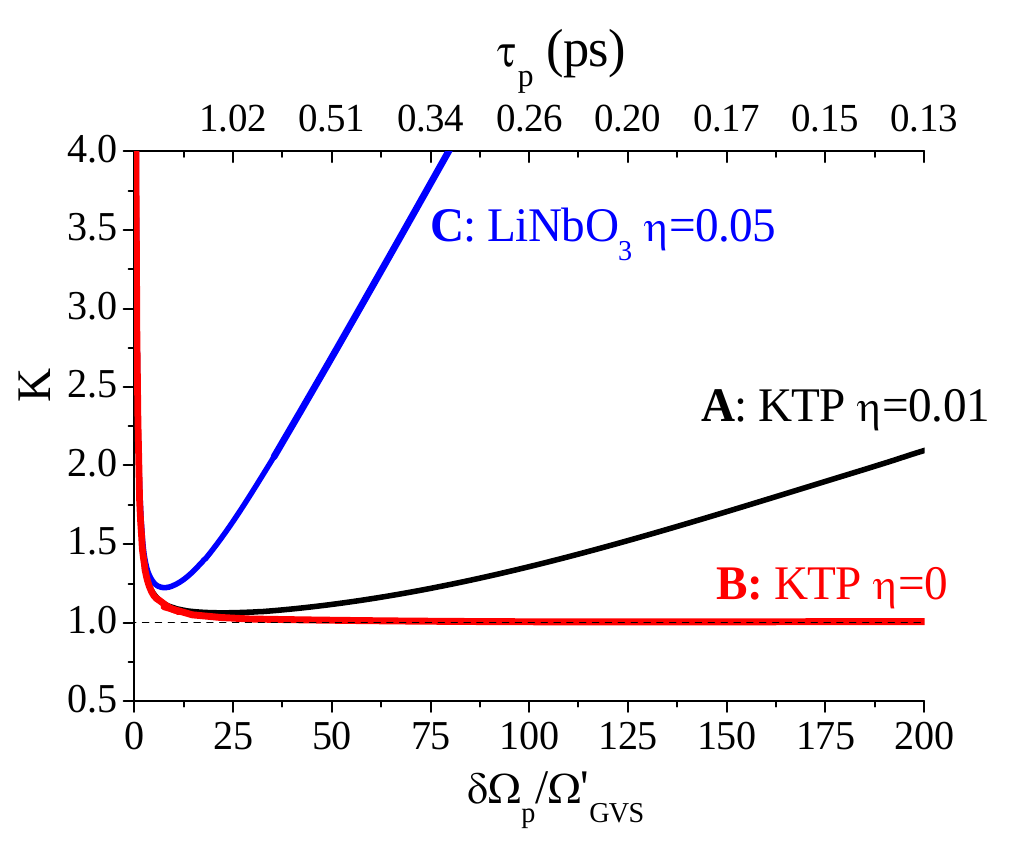}
\caption{ Role of GVM in determining the purity of the state: Schmidt number for different crystals and/or different phase matching conditions,   corresponding  to the points {\bf A}, {\bf B}, {\bf C}   in Fig.\ref{eta}:
{\bf A})  4 mm KTP  with $ \tauGVM$=0.27ps, $\tauGVS$=25.5ps, $\to \eta=0.01$  ( same  as in Figs.\ref{fig2}- \ref{kappa_001}).{\bf B}) 4 mm KTP, with 
$ \tauGVM$=0, $\tauGVS$=24.7ps, $\to \eta=0$ . {\bf C})  4 mm LiNbO$_{3} $  with  $ \tauGVM$=1.68ps, $\tauGVS$=31.2ps, $\to \eta=0.05$ }
\label{kappa_comp}
\end{center}
\end{figure}

Indeed, a  more refined calculation shows that the minimum value of $\cappa$, reached for a pump duration  intermediate between $\tauGVM$ and $\tauGVS$ is 
 $ K_{min} =1 + O(\eta)$.  This is also confirmed by analytical calculations of the Schmidt number, reported in detail elsewhere\cite{Brambilla2015},  performed by means of  a Gaussian approximation of the $\sinc$ function of  phase matching, similarly to what done in  \cite{Grice2001}.  These calculations (plotted as the blu dash-dot line in Fig.\ref{kappa_001}c)  show that the minimum of $\cappa$ is
\beq
\cappa_{min}=      \frac{1+ \eta}{1-\eta} \approx 1+ 2 \eta, 
\eeq
 reached for $\Delta \Omp= \sqrt{3 \OmGVS \OmGVM}$
This result suggest that a higher degree of purity of the reduced states can be achieved as the GVM between the two forward propagating is reduced. This is confirmed by the examples in Fig.\ref{kappa_comp},  which compares different crystals and phase matching conditions.  Notice that a small $GVM$ corresponds to a higher degree of purity, as in the copropagating case, but that in the present case the condition for separability  is much less  demanding, as it does not  require  a vanishing GVM,  but just that that  $\tauGVM$ is small compared to  sum of the inverse of group velocities $\tauGVS$, which is always verified to some extent. 

\section{Interpretation: the biphoton correlation in the temporal domain}
\label{sec:temporal}
\begin{figure*}[h,b,t]
\begin{center}
\includegraphics[scale=0.6]{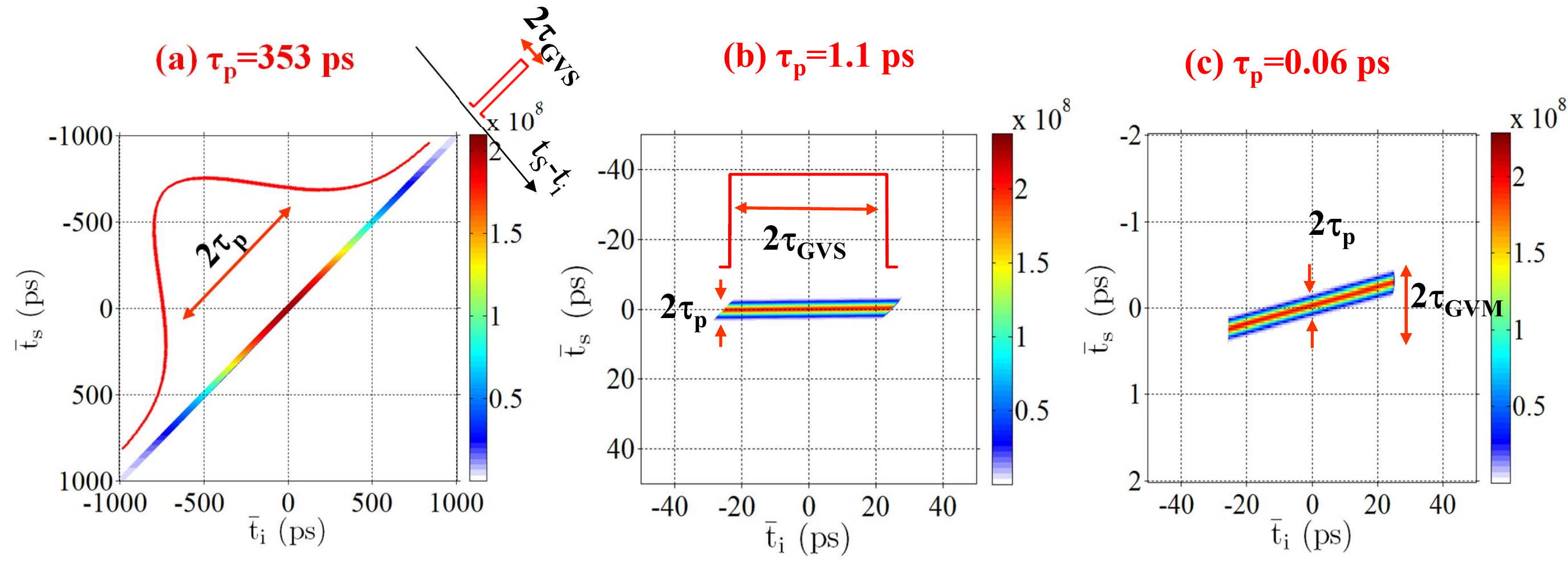}
\caption{(Color online) Temporal  correlation of twin photons  $\left|\phi (\bar t_i, \bar t _s)\right|$ , given by Eq.\eqref{phi2},    
plotted in the plane $ (\bar t_i,\bar t_s)$. (a)High  entanglement case, with $\cappa \simeq 26$ , for a quasi CW  pump $\taup  = 14 \tauGVS $.  (b)Almost separable case with $\cappa \simeq 1.06$, for an intermediate pump  $\taup  = 0.04\tauGVS=4 \tauGVM $. (c)Ultrashort pulse 
 $\taup  = 0.22 \tauGVM $, corresponding to an entangled state with $\cappa \simeq 4$. Same KTP crystal  as in Fig.\ref{kappa_001}  } 
\label{psitemporal}
\end{center}
\end{figure*}

 An alternative  insight into the issue of separability vs entanglement 
 is provided by  the analysis of the biphoton correlation in the temporal domain. 
We consider
\beq
\phi(t_s, t_i) = \langle\Asout (t_s) \Aiout(t_i)  \rangle 
= \int \frac{d\Om_s} {\sqrt{2\pi}}  \int \frac{d\Om_i} {\sqrt{2\pi}}  e^{-i (\Oms t_s + \Omi t_i)}
e^{i k_s (\Om_s) l_c }\psi(\Om_s, \Om_i) \, , 
\label{phi}
\eeq
which is proportional to the probability amplitude of finding  a signal and an idler photons at  their   crystal end faces at  times $t_s, t_i$. By using the linear approximation for phase matching \eqref{Delta1} and performing the simple Fourier transformations involved in \eqref{phi} we obtain: 
\beq
\phi(\bar t_s, \bar t_i )= \frac{g e^{ik_s l_c} }{2 \tauGVSp}  \alpha_p \left(               
 \bar t_s  + \eta \frac{ \bar t_s -\bar  t_i }{1-\eta}   \right) 
\mathrm{Rect} \left(    \frac{ \bar t_s -  \bar  t_i} { 2 \tauGVSp} \right)
\label{phi2}
\eeq
%
where
\beq
\mathrm{Rect} (x) = \left\{  \begin{array} {lcl} 1 & {\rm for} & x \epsilon \left( -\frac{1}{2}, \frac{1}{2} \right) \\
0 &  {\rm elsewhere}  & \end{array}
\right. \, ,
\label{box}
\eeq
is the box function of unitary width. The barred arguments
   $\bar t_s, \bar t_i$  denote  time intervals measured starting from  the  arrival times of  the centers of the signal/idler wavepackets.  Precisely, 
$\bar t_{s,i}= t_{s,i} -t_{As,i}$,  where
\begin{align}
 				t_{As}  &= ( k_s' + k_p ') \frac{l_c}{2}= t_{Ap} -  (k'_p- k'_s) \frac{l_c}{2}\\
 				t_{Ai}& = ( k_i' + k_p ')\frac{l_c}{2}= t_{Ap} - (k'_p- k'_i) \frac{l_c}{2}
\end{align}
where $t_{Ap} = k_p ' l_c$  is the time when   the center of the pump pulse exits the crystal slab.
Figure \ref{psitemporal} shows three examples of the temporal correlation  function \eqref{phi2}.  
\par
The general formula \eqref{phi2}  can be simplified in the limit where the pump is long with respect to $\tauGVM$, i.e.
 in the quasi CW or  intermediate limits (\ref{L1}, \ref{L3}) , where it takes the form
\beq
\phi(\bar t_s, \bar t_i) \stackrel{\tau_p \gg \tau_{GVM} }{\simeq}  g e^{ik_s l_c} \alpha_p \left(          \bar t_s    \right)
\frac{1}{2 \tauGVSp} \mathrm{Rect} \left(    \frac{ \bar  t_i  -   \bar t_s} { 2 \tauGVSp} \right)\, , 
\label{phi3}
\eeq
Indeed, when the the pump pulse is  long with respect to $\tauGVM$,  we have 
$ \alpha_p \left(     \bar t_s +   \eta \frac{ \bar t_s -\bar  t_i }{1-\eta}   \right)  \approx \alpha_p (\bar t_s)$ ,  because $|\bar t_s -\bar t_i| $  is limited by the box function to values smaller than $\tauGVSp$, 
so that $  \eta \frac{| \bar t_s -\bar  t_i|  }{1-\eta} = \frac{\tauGVM}{\tauGVSp} |\bar t_s -\bar  t_i| \le \tauGVM \ll \tau_p$ . 

Formula \eqref{phi3} shows that in the limit of a negligible $GVM$, 
the distribution of separations $ \bar t_s -\bar t_i $ between the arrival times of the twin photons is  entirely 
described by the box function of width $2\tauGVSp$. 
This form  of  the temporal correlation clearly reflects  the spontaneous character of the process, where photon pairs can be  generated at any point of the crystal with uniform probability. Thus, assuming for simplicity that the the twins travel with the same group velocities $v_{gs} =v_{gi} $,  the  separation between their  arrival times  ranges with uniform  probability from  zero, when the two photons are generated at the center of the crystal up to $\pm \tauGVSp=l_c/v_g$, when they are generated at each of  the end faces.  \footnote{Precisely,
when the two photons are generated at the crystal center
$t_s-t_i= t_{As}-t_{Ai}=( k'_s-k'_i)l_c/2 \approx 0$ ,  and 
 the delay between their  arrival times   ranges uniformly between  i) $t_s-t_i=t_{As}-t_{Ai} -\tauGVSp =- k'_i l_c $, when they are generated at the right end face of the slab,
and ii)  $t_s-t_i= t_{As}-t_{Ai} +\tauGVSp = k'_s l_c $ when the photon pair is generated at the left end face.}
\par
The  CW pump limit (Fig.\ref{psitemporal}a) corresponds  to the situation where   the pump pulse is much longer than the maximal temporal separation $\tauGVSp$  between the twins. In this case, the usual picture of the temporal entanglement of twin photons holds: the time when a signal or idler photon  is  individually detected has a large  undeterminacy, because  a photon pair can be generated at any time along the pump pulse. However,  from the arrival time of one of the members of the pair one can infer the arrival time of the other  with a much smaller  uncertainty $\tauGVSp$,  which  represents the mean uncertainty in the arrival time of one photon provided its twin  have been detected, i.e. the {\em correlation time}.  This kind of correlation is basically  what predicted in Ref. \cite{Suhara2010} for a strictly monochromatic pump. 

However, when the pump pulse shorten below $\tauGVSp$ (Fig.\ref{psitemporal}b) 
this description ceases  to be valid, because the localization of the pump pulse provides a timing information on the arrival time of the signal that is more precise than the uncertainty in the temporal separation of  the twins. 
Indeed when the pump pulse is much shorter than $\tauGVSp$, but still long enough that GVM is negligible, the signal wavepacket overlaps almost exactly with the pump pulse during propagation, and the uncertainty in the arrival time of the signal is just the pulse duration. This  is much smaller than the conditional uncentainty $\tauGVSp$ by which the arrival time of the idler can be inferred from that of the signal, so that the arrival times  of the members of the pair appear completely uncorrelated. 
Indeed, the temporal correlation in Fig.\ref{psitemporal}b  is approximately: 
\beq
\phi(\bar t_s, \bar t_i) \simeq  g e^{ik_s l_c} \alpha_p \left(          \bar t_s    \right)
\frac{1}{2 \tauGVSp} \mathrm{Rect} \left(    \frac{ \bar  t_i  } { 2 \tauGVSp} \right)\, , 
\label{phi3}
\eeq
which is a factorable function of   $\bar t_s, \bar t_i$. 

Notice that when the pump pulse is so short that GVM starts to be important (Fig.\ref{psitemporal}c), there  is again  a loss of absolute timing information,  because in this case the arrival time of the signal cannot be inferred from that  of the  pump  with a precision better than $\tauGVM$.  In contrast,  the arrival time of the signal {\em conditioned } to a photon count in the idler arm 
 can be predicted within  the short pump  duration $\tau_p$, and the state becomes again entangled. 
This can be better understood by looking  the correlation function \eqref{phi2}, which  for  $\tau_p \ll \tauGVM$ can be rewritten as 
\begin{align}
\phi(\bar t_s, \bar t_i )&= 
\frac{g e^{ik_s l_c} }{2 \tauGVSp}  \alpha_p \left(               
\frac{ \bar t_s  -  \eta \bar  t_i }{1-\eta}   \right) 
\mathrm{Rect} \left(    \frac{ \bar t_s -\bar t_i } { 2 \tauGVSp} \right) \\
& \simeq \frac{g e^{ik_s l_c} }{2 \tauGVSp}  \alpha_p \left(               
\frac{ \bar t_s  -  \eta \bar  t_i }{1-\eta}   \right) 
\mathrm{Rect} \left(    \frac{ \bar t_s } { 2 \tauGVM} \right) \, .
\label{phi4}
\end{align}
where the last line has been obtained by substituting $\bar t_i = \bar t_s /\eta$ inside the argument of the box function (valid because the pump profile is much narrower than both $\tauGVSp$ and $\tauGVM$).
From  formula \eqref{phi4} we see that,   provided that an idler photon is detected, say at time $\bar t_i $, the arrival time of the signal can be predicted 
as $\bar t_s = \eta \bar t_i $
within the narrow uncertainty of the pump duration $\tau_p$  (see also Fig.\ref{psitemporal}c). 
 However when the idler is not detected, the overall uncertainty in the signal arrival time is the larger width  $\tauGVM$  of the box function. Clearly this argument predicts an entangled state, with the  number of modes scaling  as $\tauGVM/\tau_p$, in agreement with 
formula \eqref{kappaultrashort}.

%
\section{Conclusions}
In this work we provided a detailed theoretical analysis of the effect of the pump spectral properties on the quantum correlation of counterpropagating photons generated by SPDC in a periodically poled crystal. 

 In particular,  for increasing  spectral bandwidths of the pump (descreasing pump durations),  we demontrated a remarkable transition from a high-dimensional entangled state, to an almost separable state.  The transition occurs when the pulse duration shorten below 
the characteristic  transit time $\tauGVSp = \frac{l_c} {2v_{gs} } + \frac{l_c} {2v_{gi} }\approx \frac{l_c} {2v_{gp} }+ \frac{l_c} {2v_{gi}}$
of light along the crystal slab.  This {\em long}  temporal scale  is a unique characteristic of the counterpropagating geometry, being associated to the  delay between the times at which  the  counterpropagating  photons, generated at some point along the slab,   appear at their exit  faces.  
The temporal correlation (temporal entanglement) is again restored for  pump durations below the {\em short}  temporal delay  occurring between the co-propagating waves because of their different group velocities. \\
The natural existence of such  separated  time scales ensures the possibility of generating 
high purity single photon ( i.e. a separable two-photon state), under very general conditions, which differs drastically from the usual co-propagating geometry \cite{Mosley2008}. 
\par
These conclusions have been  supported throught the paper by the analysis of the Schmidt number in Sec.\ref{sec:Schmidt}, and by  analytical  and numerical evaluations of the spectral and temporal correlation function (Sec.\ref{sec:psi},  Sec.\ref{sec:temporal}).
\par
The study of the maginal statistics of photons  in Sec. \ref{sec:coherence} has revealed several non-trivial features: \\
While for a long pump pulse twin photons have the same spectrum and the same  coherence properties, in the regime of separability they  exhibit very different features.  In particular, the properties of the counterpropagating idler are entirely  determined by the phase matching in the medium,  so that we can say that they  reflect the momentum conservation in the process. 
On the other side, the spectro-temporal properties of the signal are a replica of those of the co-propagating  pump laser pulse, and rather reflect the energy conservation. \\
 For an ultrashort pump pulse, our quantum analysis has  retrieved  results analogue  to  what  predicted  in the  classical description of the MOPO \cite{Canalias2007, Stromvqist2012}, but with some additional limitation.   At difference with the MOPO prediction   \cite{Canalias2007},   our  results impose a precise inferior limit to the observable  bandwidth of the backward idler  photon, which cannot be narrower than the phase matching bandwidth $\OmGVSp$.   Clearly, our analysis is  limited to SPDC, but we notice that, to our knowledge, measurements of the spectrum of the backward  wave 
in the MOPO  have been  limited by the spectrometer resolution \cite{Pasiskevicius2015},  so that  
our findings may open a 
question about the effective bandwidth of the backward wave.  
\label{sec:conclusions}
\appendix
\section{ Approximations for the biphoton amplitude}
\label{A1}
 In this Appendix  we derive the approximated forms of the biphoton amplitude used in the text, which holds  in the  various  pump regimes.  In all the cases we make use of  the linear approximation for phase matching  \eqref{Delta1}, based on the assumption that the bandwidths in play are narrow so that dispersion can be neglected. 
Under this approximation,  the  general espression \eqref{psi} of the biphoton amplitude becomes
\beq
 \psi (\Oms, \Omi)  = \frac{g}{\sqrt{2\pi}}   
 \aptilde \left( \Oms +\Omi \right)  V \left( \frac{\Oms}{\OmGVM}  + \frac{\Omi}{\OmGVS}  \right)  
\label{A1} 
\eeq
where for brevity of notation we introduced  the phase matching function 
$ V(s)  = 
 \sinc \left( s  \right)     e^{i s }   $. 
\par
We consider first  the limit of a CW pump \eqref{L1}. Since the pump bandwith is much narrower than  the bandwidths $\OmGVS$ and $\OmGVM$ of phase matching,   the presence of the pump Fourier amplitude term forces $\Oms = -\Omi $ into the phase matching function. As a result 
\begin{align}
\lim_{{\tau_p}/{\tauGVS} \to \infty}  \psi (\Oms, \Omi) & = \frac{g}{\sqrt{2\pi}}    \aptilde \left( \Oms +\Omi \right) 
V \left( -\frac{\Oms}{\OmGVSp}  \right)   \\
& = \frac{g}{\sqrt{2\pi}}    \aptilde \left( \Oms +\Omi \right) 
V \left( \frac{\Omi}{\OmGVSp}  \right)   
\end{align}
where  we used the relation 
$1/ \OmGVSp  = 1/ \OmGVS  -1/\OmGVM  $, according to the definitions (\ref{GVM} -\ref{GVSp}). 
\par 
The limit \eqref{L2} of an ultrashort pump is also straightforward. In this case the bandwidths of phase matching are assumed to be much narrower than the pump bandwidth $\OmGVS \ll \OmGVM \ll \DOmp$, so that the phase matching function has a narrow peak, which on the slow scale of variation of the pump forces  $\Omi  = - \eta \Oms  $,  or $\Oms = - \Omi /\eta$ inside the pump argument. Therefore 
\begin{align}
\lim_{\frac{\tau_p}{\tauGVM} \to 0}  \psi (\Oms, \Omi) & = \frac{g}{\sqrt{2\pi}}    \aptilde \left[ \Oms (1-\eta) \right] 
V \left( \frac{\Oms}{\OmGVM}  + \frac{\Omi}{\OmGVS}  \right)  
   \\
& = \frac{g}{\sqrt{2\pi}}     \aptilde \left[ -\Omi \frac {1-\eta}{\eta}  \right] 
 V \left( \frac{\Oms}{\OmGVM}  + \frac{\Omi}{\OmGVS}  \right)  
\end{align}

The intermediate pump limit \eqref{L3}  where $ \OmGVS \ll \DOmp \ll \OmGVM$ is a bit more involved.  We remind that the existence of this limit also requires $\eta = \OmGVS/\OmGVM \ll1$, which is in practice always verified to some extent. 
By introducing the pump frequency  $\Omp=     \Oms + \Omi$ , we recast the argument of the $\sinc$  function
\beq
 \frac{\Oms}{\OmGVM}  + \frac{\Omi}{\OmGVS}  =  \frac{\Omp }{\OmGVM}  + 
\Omi \left( \frac{1 }{\OmGVS}  - \frac{1 }{\OmGVM}\right)
\approx  \frac{\Omi}{\OmGVSp}     \label{A3} 
\eeq
where  the term $ \Omp / \OmGVM  $  has been neglected because is on the order $ \DOmp / \OmGVM \ll 1 $ \\
Concerning  the pump amplitude we recast it as: 
\beqa
 \aptilde  \left( \Oms +\Omi \right)& =&
 \aptilde \left[ \Oms \left( 1  -\eta\right)
 + \left( \frac{\Oms}{\OmGVM}    +  \frac{\Omi}{\OmGVS}    \right)  \OmGVS \right]  \label{A5} \nn \\
&\approx & \aptilde \left[ \Oms \left( 1  -\eta\right) \right] 
\label{A6} 
\eeqa
where the approximation in the second second line holds because   
$\left( \frac{\Oms}{\OmGVM}  + \frac{\Omi}{\OmGVS}  \right) $  is  the argument of the $\sinc$ function (see  Eq.\eqref{A1}),  so that 
it is  limited to values inside  the bandwidth of the $\sinc$, say on the order $\simeq 10$  .  Provided thar  $\OmGVS/ \DOmp $  is small enough,   this term becomes therefore negligible. 
With this in mind we can write the limiting behaviour of the biphoton amplitude: 
\begin{flalign}
 \lim_{\stackrel{\tau_p /\tauGVS \to 0}{\tauGVM/\taup \to 0 }}  
\psi (\Oms, \Omi) =&  \frac{g }{ \sqrt{2\pi} }     \aptilde \left[ \Oms \left( 1  -\eta \right) \right]    e^{i \frac{\Oms}{\OmGVM}  }  
\times 
\sinc \left(  \frac{\Omi}{\OmGVSp}  \right)   e^{i  \frac{\Omi}{\OmGVSp}  }      \label{psiL3} \\
\approx& 
\frac{g }{ \sqrt{2\pi} }     \aptilde \left[(\Oms \right)    e^{i \frac{\Oms}{\OmGVM}  }  
\times 
\sinc \left(  \frac{\Omi}{\OmGVSp}  \right)   e^{i  \frac{\Omi}{\OmGVSp}  }    \, .  \label{psiL3b} 
\end{flalign}
where the approximation in the last line is not mandatory, but  could be useful in order to get consistent results, because clearly this limit can be realized only for  $\eta = \tauGVM/\tauGVS \to 0$. 

\end{document}